\documentclass[amsfonts, amssymb, amsmath, subscriptaddress, super, showpacs, reprint]{revtex4-1}
\usepackage{bm}
\usepackage{hyperref}
\usepackage{graphicx}
\usepackage{epsfig} 
\usepackage{color}

\def\ie{\mbox{\em i.e.\ }}

\def\kk{\mathbf k}

\begin{document}

\title{The anisotropic redistribution of free energy for gyrokinetic plasma turbulence in a Z-pinch}
\author{Alejandro \surname{Ba\~n\'on Navarro}}
\email{banon@physics.ucla.edu}
\affiliation{Department of Physics and Astronomy, UCLA, 475 Portola Plaza, Los Angeles, CA 90095-1547, USA}
\author{Bogdan Teaca}
\email{bogdan.teaca@coventry.ac.uk}
\affiliation{Applied Mathematics Research Centre, Coventry University, Coventry CV1 5FB, United Kingdom}
\affiliation{Max-Planck/Princeton Center for Plasma Physics}
\author{Frank Jenko}
\email{jenko@physics.ucla.edu}
\affiliation{Department of Physics and Astronomy, UCLA, 475 Portola Plaza, Los Angeles, CA 90095-1547, USA}
\affiliation{Max-Planck/Princeton Center for Plasma Physics}
\begin{abstract}
For a Z-pinch geometry, we report on the nonlinear redistribution of free energy across scales perpendicular to the magnetic guide field, for a turbulent plasma described in the framework of gyrokinetics. The analysis is performed using a local flux-surface approximation, in a regime dominated by electrostatic fluctuations driven by the entropy mode, with both ion and electron species being treated kinetically. To explore the anisotropic nature of the free energy redistribution caused by the emergence of zonal flows, we use a polar coordinate representation for the field-perpendicular directions and define an angular density for the scale flux. Positive values for the classically defined (angle integrated) scale flux, which denote a direct energy cascade, are shown to be also composed of negative angular sections, a fact that impacts our understanding of the backscatter of energy and the way in which it enters the modeling of sub-grid scales for turbulence. A definition for the flux of free energy across each perpendicular direction is introduced as well, which shows that the redistribution of energy in the presence of zonal flows is highly anisotropic.
\end{abstract}
%
%
\maketitle

\section{Introduction} 
In plasma physics, turbulence is an underlying  problem for numerous topics of research, ranging from the study of anomalous transport across magnetic surfaces in fusion relevant plasmas to the problem of plasma heating in astrophysical conditions. One of the most identifiable characteristics of turbulence is the self-organization of turbulent structures. Mathematically, following a Fourier decomposition, the nonlinear interactions between various scales of the system can be discerned (here length scales being defined simply as $\!1/k$) and the self-organization of turbulence can be described in term of energy transfers and scale fluxes\cite{Kraichnan:1959p1578}. While this methodology is primarily used in the case of homogenous isotropic turbulence, large scale motions like the zonal flows are captured by particular modes of the system. As these modes can lead to the development of an anisotropy that can affect the energy transfers and the scale fluxes, expanding the definitions of various nonlinear based diagnostics becomes necessary for a better understanding of the underlying dynamics of the turbulent system and the subsequent development of models.    

Micro-turbulence in gradient driven magnetized plasmas represents one of the best examples where large structures in turbulence self-organize to give rise to zonal flows\cite{Diamond:2005p222,plunk15} that in turn dominate the self-organization of smaller scales. Since the turbulent structures in plasma reach scales much smaller than that of the ion gyroradius in the presence of a strong magnetic guide field, using the five-dimensional gyrokinetic (GK) formalism (for which, due to the gyrokinetic ordering, the gyration phase of charged particles around the magnetic guide field has been removed consistently\cite{Brizard:2007p11}) represents the most efficient method for the description of the problem\cite{Krommes:2012p1373}. 

This approach\cite{Villard:2010p1254} is of particular interest in sheared toroidal magnetic geometries characteristic of tokamaks. As turbulent structures tend to elongate along magnetic field lines, the natural coordinate system for the problem, namely field-aligned coordinates\cite{Beer:1995p1292} $(x,y,z)$, becomes nonorthogonal. Any modal decomposition along nonorthogonal directions show complicated couplings of primary instabilities (responsible for the source of free energy; the quantity of interest for the nonlinear redistribution\cite{Schekochihin:2008p1034} in GK turbulence) and secondary instabilities (responsible for the self-organization of structures). To simplify the analysis of the interplay between primary and zonal-flow modes and the subsequent impact on the turbulent energy redistribution, we turn towards a simpler yet non-trivial problem, that of the Z-pinch\cite{Haines:2011p1664}.

In a Z-pinch configuration, described using a simple cylindrical coordinate system ($\tilde r,\tilde \theta,\tilde z$), a cylindrically symmetric equilibrium magnetic guide field is given as, ${\bf B}=B_0(\tilde r) {\bf{e}_{ {\tilde \theta}}}$. Unlike the complicated tokamak geometry (even local equilibrium models\cite{Lapillonne:2009p1355} have a non-diagonal metric tensor), for such a simple closed field line system a Fourier mode can only contribute energy to one scale, drastically simplifying the energy redistribution analysis. The primary instability driving GK turbulence in this system is the small scale entropy-mode\cite{Ricci:2006p1665}. It manifests itself by the development of streamer like structures in the radial direction. As secondary instabilities are excited, zonal flows develop in the $\tilde z$-direction. Depending on the parameters of the system, zonal flows will dominate the self-organization of turbulence\cite{Kobayashi:2009p1091, Kobayashi:2012p1656}, leading to a predator-prey type oscillation\cite{Kobayashi:2015p1677} between streamer and zonal flows like structures. Form this interplay, coherent vortices can emerge, as seen in other plasma configurations\cite{Nakata:2010p1678}. 

While the study of Z-pinch plasmas represents an area of interest on its own, we limit our analysis to the impact made by zonal flow structures on the redistribution of energy between scales and the subsequent wavenumber anisotropy developed. As such, we introduce a series of nonlinear diagnostics that can describe the wavenumber anisotropy from the perspective of free energy fluxes. This represents the focus of the current work, for which the Z-pinch configuration is used as a simple framework for these ideas to be put forward. In the final section, we will discuss an extension of these diagnostics and their applicability to tokamak geometries.

\section{The Z-pinch gyrokinetic system} 

Numerically, we make use of the Eulerian code {\sc Gene}\cite{Jenko:2000p1248}, which can solve the nonlinear gyrokinetic equations in both global\cite{Gorler:2011p1340} and local\cite{Merz2009} (flux-tube) geometries. Here, the standard field-aligned coordinates $(x,y,z)$ are employed for a local flux-surface approximation of the Z-pinch geometry ($\tilde r,\tilde \theta,\tilde z$). In this approximation, a small periodic domain, mapped by the radial coordinate $x$, is centered around a specific flux-surface identified by $\tilde r=R$. As the equilibrium magnetic field lines are hard-coded along the third direction ($z=\tilde \theta$), the $y$ coordinate represents the mapping of $- \tilde z$. The macroscopic normalization length, used throughout our work to normalize gradient lengths and curvature terms for example, is set to $R$. With this in mind, performing a Z-pinch geometry run in {\sc Gene} requires no special alteration to the code. It is only necessary to provide the appropriately normalized cylindrical metric coefficients ($g^{xx}=g^{yy}=g^{zz}=1$, $g^{xy}=g^{xz}=g^{yz}=0$), the resulting Jacobian ($\sqrt g=1$) and the magnetic radial gradient $K_y \equiv \frac{R}{B_0}\frac{d B_0}{d x}= -1$ (using Ref.~[\onlinecite{Lapillonne:2009p1355}] notations for the magnetic curvature coefficients; listing only the nonzero contribution). The velocity space coordinates ${ \{v_{\parallel}, \mu \} }$ are, respectively, the velocity parallel to the magnetic field and the magnetic moment (containing the perpendicular velocity information). 

The system is near a state of equilibrium, prescribed for the species $s$ by an appropriately normalized Maxwellian distribution, $F_{0s}=\pi^{-3/2}e^{-(v_\parallel^2+\mu B_0)}$. We use a kinetic description for both ion and electron species, with the mass ratio $m_i/m_e=1836$ and equal background temperature ($T_{0i}=T_{0e}$) and density ($n_{0i}=n_{0e}$; quasi-neutrality). The coordinates  perpendicular to the magnetic field are Fourier transformed $(x,y) \rightarrow (k_x,k_y)$ and the GK system is solved numerically using respectively $\{N_{k_x},N_{k_y}, N_z, N_{v_\parallel}, N_{\mu}\}$=$\{128, 64, 16, 64, 32\}$ modes/points in each direction. In the perpendicular directions, the smallest wavenumbers considered are $k^{\min}_x=k^{\min}_y=0.1$. Due to the reality condition, only positive $k_y$ modes are solved, the largest wavenumbers considered being equal ($k_x^{\max}=k_y^{\max}=6.4$). The wavenumbers are taken in units  of the ion gyroradius ($\rho_i=v_{Ti}/\Omega_i$), the real space domain spanning $64 \, \rho_i$ units, where $v_{Ti}=\sqrt{T_{0i}/m_i}$ is the ion  thermal  velocity, $\Omega_i=q_iB_0/(m_i \mbox{c})$ is the ion cyclotron frequency and  $q_i$ is the electric charge sign for an ion species  ($q_i=1$  in units of $|e|$). The same definitions are employed for the electron species, for which naturally, $q_e=-1$. Since we limit our analysis to plasmas for which the magnetic pressure is dominant over the thermal pressure ($\beta= {8\pi n_{0i}T_{0i}}/{B_0^2} \ll 1$), we neglect additional magnetic fluctuations and consider solely electrostatic field fluctuations. For this case, from the perturbed gyro-center distribution functions $f_{s} = f_{s}(k_x, k_y, z, v_\parallel, \mu, t)$ we extract the non-adiabatic part as $h_s=f_s + q_s \frac{F_{0s}}{T_{0s}}\phi$, where $\phi$ is the gyro-averaged self-consistent electrostatic field contribution. The gyro-averaged electrostatic potential ($\phi$) is obtained in the Fourier representations simply as the Bessel function ($J_0$) screened self-consistent electrostatic potential ($\varphi$), i.e. $\phi(\kk_\perp, z)=J_0(\lambda_s) \varphi(\kk_\perp, z)$, with $\lambda_s=\sqrt{\mu B_0}|\kk_\perp| v_{Ts}/\Omega_s$. Neglecting Debye length corrections, $\varphi$ is found as, 
\begin{align}
\varphi=\frac{\pi B_0}{\sum_{s} \frac{q_s^2}{T_{0s}}n_{0s} [{1-\Gamma_0(b_s)]}} \sum_{s} n_{0s} q_s\int \!\!J_0(\lambda_s) f_s {d}v_{\parallel} {d}\mu \;,
\end{align}
with $\Gamma_0(b_s)=I(b_s)e^{-b_s}$, $I(b_s)$ the modified Bessel function and $b_s=v^2_{Ts}|\kk_\perp|^2/(2\Omega^2_s)$.

The normalized GK equation for each species s is given as,
\begin{align}
\frac{\partial f_{s}}{\partial t} = &-\left[\omega_{ns}+\left( v_\parallel^2+\mu B_0-\frac{3}{2}\right) \omega_{Ts}\right] F_{0s} ik_y\phi \nonumber\\
&-\frac{v_{Ts}}{\sqrt{g}B_0}v_\parallel \frac{\partial h_s}{\partial z} -\frac{T_{0s}(2v_\parallel^2+\mu B_0)}{q_s B_0} K_y ik_y h_s \nonumber\\
& +{\cal N}_s+{\cal H}_s+{\cal C}_s \;.
\end{align} 
where $\omega_{ns}=-\frac{R}{n_{0s}}\frac{dn_{0s}}{dx}$ and $\omega_{Ts}=-\frac{R}{T_{0s}}\frac{dT_{0s}}{dx}$ are, respectively, the normalized density and temperature gradients. Omitting the directions of no immediate interest,  ${\cal N}_s(\kk_\perp)$ is the ${\bf E}\times {\bf B}$ drift velocity nonlinearity and has the form,
\begin{align}
{ \cal N}_s(\kk_\perp)= \sum_{\kk'_\perp} [k'_xk_y-k'_yk_x]\, \phi(\kk'_\perp)h_s(\kk_\perp-\kk'_\perp) \label{eq_NL}\;. 
\end{align}
Numerically it is de-aliased using a three-halves rule. In addition, ${\cal H}_s$ refers to hyperdiffusion  terms in the  $z$ and $v_{\parallel}$  directions given by,
\begin{align}
{\cal H}_s = -\left( a_z \frac{\partial^4 }{\partial z^4} + a_{v_{\parallel}} \frac{\partial^4}{\partial v_{\parallel} ^4} \right)  h_s.
\end{align}
These hyperdiffusion terms are necessary to stabilize spurious grid-size oscillations~\cite{pueschel10}. For all the cases, the coefficients are set to  $a_z= 1.0$ and   $a_{v_{\parallel}} =0.5$, since these values have been found to be well suited for  a wide range of cases. The ${\cal C}_s$ term is a linearized Landau- Boltzmann collision operator with energy and momentum conserving terms~\cite{hauke13}, where an electron-ion collision frequency~\cite{hinton76}  of value $\nu_{ei} = 0.086$ is used in this work.  Finally, to remove any unphysical accumulation of energy at high wavenumbers, a Large-Eddy-Simulation (LES) model is used; see Ref.~[\onlinecite{BanonNavarro:2014p1535}] and references within for details regarding using LES models in GK plasmas. As it is mentioned in Ref~[\onlinecite{Kobayashi:2015p1677}], the collisional and hyperdiffusion terms can affect the growth rate of the modes. However, since this work concentrates on the study of the anisotropic redistribution of free energy, a high $k$ energy accumulation is deemed more problematic than a shift in growth rates. 

\begin{figure}[htb]
\begin{center}
\includegraphics[width = 0.48\textwidth]{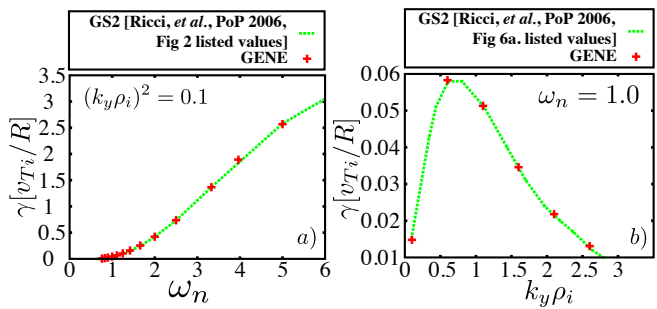}
\end{center}
\caption{(Color online) Benchmark of the linear growth rate for the entropy mode as a function of a) $\omega_n$ and b) $k_y$.}
\label{fig_benchmark}
\end{figure}

As numerically $\beta$ is taken to be zero, the pressure gradient is neglected and the ideal MHD interchange mode does not enter into our system. This is equivalent to the ideally stable parameter regime in which the free energy injected in the system is due only to the entropy-mode instability\cite{Ricci:2006p1665, Kobayashi:2012p1656}. In Fig.~\ref{fig_benchmark}, as this is the first time {\sc Gene} has been used for a Z-pinch geometry, the linear growth rates for the entropy mode in the collisionless case are benchmarked against the GS2 values that are listed by Ref.~[\onlinecite{Ricci:2006p1665}] in Figs.~2 and 6a, therein.

Throughout this work we take $\omega_{Ts}=0$ and consider two values for $\omega_{ns}= \{1.6, 4.0\}$. The $\omega_{ns}= 1.6$ case has been used and reported on in other works\cite{Ricci:2006p1665, Kobayashi:2015p1677} in the literature.

\section{Energy conservation and fluxes}

\subsection{Scale flux}

The analysis of the nonlinear energy redistribution between perpendicular spatial scales, presented here, is based solely on the computation of ${\cal T}_s(k_x, k_y)$, defined as
\begin{align}
{\cal T}_s(k_x, k_y)\!= \!\int \!\! \frac{T_{0s}}{2F_{0s}}\Re\{ { \cal N}_s(k_x,k_y) h_s^*(k_x,k_y)\} \, dz dv_\parallel d\mu. 
\label{eqT}
\end{align} 
From the perspective of the underlying triadic interactions, ${\cal T}_s(k_x, k_y)$ can be interpreted as the energy received by a mode $\bf k_\perp$ from the interaction with all other modes. Seeing ${\cal T}_s(k_x, k_y)$ as the total energy received rather than the total energy given is just an arbitrary choice that does not change the results as long as this interpretation is used consistently; here a mode receives energy if ${\cal T}_s>0$ and gives energy if ${\cal T}_s<0$. Unlike previous works\cite{BanonNavarro:2011p1274, BanonNavarro:2014p1535, Teaca:2012p1415, Teaca:2014p1571}, which use a shell decomposition for the nonlinear energy transfers that requires multiple computations of the nonlinear terms (of the order of the number of shells, or even its square), an analysis based solely on ${\cal T}_s$ terms requires only one computation of the nonlinear term. While using this approach prohibits the computation of shell-to-shell energy transfers\cite{BanonNavarro:2011p1274} or obtaining information related to the locality of interactions\cite{Teaca:2012p1415, Teaca:2014p1571}, it is still possible to obtain data related to the transfer spectra and the energy fluxes, as we will see next.  

The conservation of energy by the nonlinear interactions implies that  
\begin{align}
\int_{-\infty}^{+\infty} \int_{0}^{+\infty} {\cal T}_s(k_x, k_y) \, dk_x dk_y=0 \;,
\label{sumTxy_zero}
\end{align} 
where the second integral limit is taken from $0$ rather than $-\infty$ due to the reality condition for the dynamical quantities, which translates to ${\cal T}_s(k_x, k_y)={\cal T}_s(-k_x, -k_y)$. The free energy is independently conserved by the ${\bf E}\times {\bf B}$ nonlinear term for each species. However, it would be improper to think of the nonlinear interactions that contribute to these two individual energy channels as being separate. As the gyro-averaged electrostatic potential enters the nonlinearities from the ${\bf E}\times {\bf B}$ drift velocity, it acts as a mediator for the energy exchanges. Seeing the electrostatic potential as the sum of moments of the distribution functions, leads to the interpretation that each species mediates the energy exchanges of all other species, or equivalently that each energy transfer for a species is mediated by all the other species. 

Rewriting the condition given by Eq.~(\ref{sumTxy_zero}) in term of a polar decomposition, \ie $k_x=k \cos(\theta), k_y=k \sin(\theta)$, allows for a more transparent interpretation in terms of scales ($k\equiv |\bf k_\perp|$),  
\begin{align}
\int_{0}^{+\infty} \int_{0}^{\pi} {\cal T}_s(k, \theta) \, kdk d\theta=0 \;,
\label{sumTktheta_zero}
\end{align} 
where again we use the fact that the reality condition implies ${\cal T}_s(k, \theta)= {\cal T}_s(k, \theta\pm\pi)$. The transfer spectra ${\cal T}_s(k)$ is then simply defined as
\begin{align}
{\cal T}_s(k)= \int_{0}^{\pi} k{\cal T}_s(k, \theta) \, d\theta \;.
\end{align}
Since the integral of $\mathcal T_s(k)$ across all $k$'s is zero, splitting this integral in regard to a cutoff wavenumber $k_c$, 
\begin{align}
\int_{0}^{k_c} \mathcal T_s(k) dk+\int_{k_c}^{+\infty} \mathcal T_s(k) dk= 0\;,
\label{tt0}
\end{align} 
leads to two terms of equal value and opposite sign. This forms the basis for the definition of the energy flux across a scale $k_c$, 
\begin{align}
\Pi_s(k_c)&=\int_{k_c}^{+\infty} \mathcal T_s(k) dk=-\int_{0}^{k_c} \mathcal T_s(k) dk \nonumber \\
&=\frac{1}{2} \bigg{[}\int_{k_c}^{+\infty} \mathcal T_s(k) dk -\int_{0}^{k_c} \mathcal T_s(k) dk \bigg{]}\;.
\label{def_flux}
\end{align} 
This definition takes into account that $\mathcal T_s(k)$ stands in for the net energy received by a scale $k$ and leads to a positive value flux for a flow of energy from large to small scales (direct cascade). Consequently, a negative value for the flux denotes an inverse cascade of energy across scales.

\subsection{Angular flux}

In a similar fashion, we can define an angular flux that measures how much energy is flowing across a direction that cuts across all $k$'s, identified by an angle $\theta_c$. Considering that the quantity 
\begin{align}
\sigma_s(\theta)=\int_{0}^{\infty} k{\cal T}_s(k, \theta) \, dk \;,
\end{align}
is invariant under a $\pi$ rotation, \ie $\sigma_s(\theta)=\sigma_s(\theta \pm \pi)$, the integral of $\sigma_s(\theta)$ over any $\pi$ interval is zero. Shifting the integration limits by $\pm \pi/2$ around the direction of interest and splitting the integral with respect to $\theta_c$ leads to two terms of equal value and opposite sign, 
\begin{align}
\int_{\theta_c-\frac{\pi}{2}}^{\theta_c} \sigma_s(\theta) d\theta+\int_{\theta_c}^{\theta_c+\frac{\pi}{2}} \sigma_s(\theta) d\theta=0 \;.
\end{align} 
Doing so ensures that the two terms sweep over the same number of modes regardless of the $\theta_c$ choice, simplifying the interpretation. The angular flux $\Xi(\theta_c)$, taken as positive if it occurs in a trigonometric direction, is now defined as, 
\begin{align}
\Xi(\theta_c)=\frac{1}{2} \bigg{[}\int_{\theta_c}^{\theta_c+\frac{\pi}{2}} \sigma_s(\theta) d\theta -\int_{\theta_c-\frac{\pi}{2}}^{\theta_c} \sigma_s(\theta) d\theta \bigg{]}\;.
\label{def_angflux}
\end{align} 

An ideally isotropic system will have a zero angular flux across all directions. It should be noted that the angular flux can be defined to be positive if energy is being moved towards a given direction (towards the $k_y$ direction for instance). While simpler conceptually, the definition given here leads a transfer of energy from the $k_x$-axis towards the $k_y$-axis to generate a positive flux in the $\theta_c \in[0, \pi/2]$ interval and a negative value one in the $\theta_c \in [\pi/2, \pi]$ interval.

\subsection{Angular density of the scale flux} 

We introduce another quantity that can help characterize the anisotropic redistribution of free energy. The angular density of the free energy scale flux can be defined as, 
\begin{align}
\Pi_s(k_c, \theta)=\int_{k_c}^{+\infty}  {\cal T}_s(k, \theta) \, kdk  \;.
\label{def2_fa}
\end{align} 
From the start it is evident that the scale flux is recovered by integrating over $\theta$,
\begin{align}
\Pi_s(k_c)=\int_0^{\pi}\Pi_s(k_c, \theta)d\theta \;,
\label{def2_f}
\end{align} 
highlighting the role of $\Pi_s(k_c, \theta)$ as an angular density of the free energy scale flux. However, while the three definitions of the scale flux given by Eq.~(\ref{def_flux}) are equivalent, the resulting angular densities are not. The choice adopted here, is dictated by practical concerns. Numerically, we obtain ${\cal T}_s(k, \theta)$ from an angular decomposition of  ${\cal T}_s(k_x, k_y)$, using here angular sections spawning $3^\circ$ arcs each. As small $k$ angular sections can fail to capture any modes from the rectangular grid, it is preferred to use the given definition (computed numerically in the interval $[k_c, k^{\max}]$), which is not as sensitive to the angular distribution of small $k$ modes. While the results obtained from the definition adopted here are smoother in appearance, the visual representation of the anisotropy is the same regardless of the definition adopted.

We stress that the angular density of the scale flux is still a quantity related to the transfer of energy across scales and it should not be interpreted as a measure of the movement of energy from one direction to another (which is the angular flux given by Eq.~\ref{def_angflux}).

\section{The results}

\subsection{A description of the turbulent states}

\begin{figure}[tb]
\begin{center}
\includegraphics[width = 0.48\textwidth]{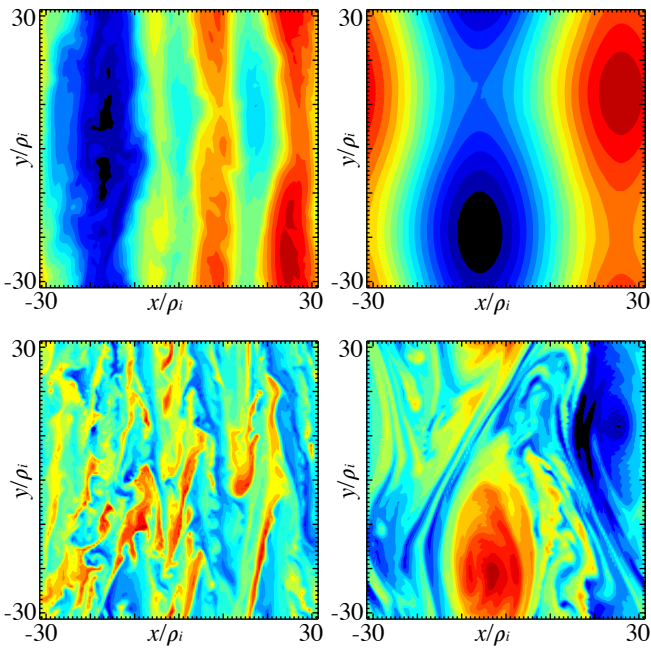}
\end{center}
\caption{(Color online) Representation of the typical (top) electrostatic potential $\phi$ and (bottom) density fluctuations in the steady state for (left) $\omega_{ns}= 1.6$ and (right) $\omega_{ns}= 4$ cases. The fields are integrated over the $z$ direction.}
\label{fig_phi}
\end{figure}

\begin{figure}[b]
\begin{center}
\includegraphics[width = 0.48\textwidth]{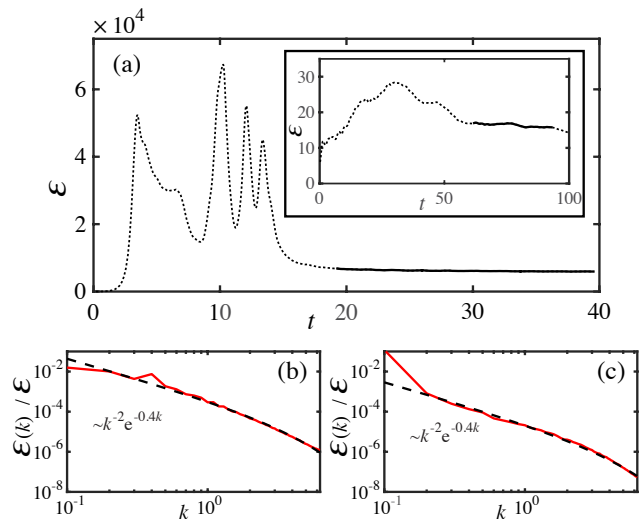}
\end{center}
\caption{(Color online) (a) The free-energy evolution in time for $\omega_{ns}= 4$; the insert shows the same quantity for $\omega_{ns}= 1.6$. For both cases, the full black line represents the averaging interval used for steady state diagnostics. The averaged free-energy spectra, normalized to the averaged global free energy values, in the steady state for b) $\omega_{ns}= 1.6$ and c) $\omega_{ns}= 4$. The dashed black line fits the listed theoretically scaling.}
\label{fig_free-energy}
\end{figure}

We report on the steady state regimes for the $\omega_{ns}= 1.6$ and $\omega_{ns}= 4$ cases; the gradient levels for both ion and electron species are equal. In both cases, zonal flows develop that shear the radially ($x$) elongated structures (streamers). The steady state for $\omega_{ns}= 1.6$ is characterized by elongated structures in the $y-$direction for the gyro-averaged electrostatic potential, Fig.~\ref{fig_phi}. This case represents a textbook example of what we come to associate with zonal flows.

However, the steady state for the $\omega_{ns}= 4$ case presents a series of peculiarities. A recent work\cite{Kobayashi:2015p1677} talks about the predator-prey behavior of zonal flows (predator) and turbulent streamers (prey). The case $\omega_{ns}= 4$ exhibits such a behavior during its transient state, consisting of streamer like structures being broken by zonal flows only to reform again. This interplay during the transient state corresponds in Fig.~\ref{fig_free-energy}-(a) to the large oscillations seen in the time evolution of the free energy ($ {\cal E}\!=\sum_s \!\int \!\! \frac{T_{0s}}{2F_{0s}}f_s h_s \, dx dy dz dv_\parallel d\mu$). As this behavior subsides, coherent vorticies\cite{Nakata:2010p1678} emerge as the dominant structures during the steady state regime, as seen in Fig.~\ref{fig_phi}. 

For $\omega_{ns}= 4$, a lot of the energy is contained at seemingly frozen large scales. However, smaller scales are energetically active, as seen from the free energy spectra, Fig.~\ref{fig_free-energy}-(c). For both cases, due to the nonlocal influence of the zonal flows, the free energy spectra seems to follow a $\sim k^{-2}e^{-0.4k}$ law, consistent with the prediction made by Ref.~[\onlinecite{Kobayashi:2015p1677}] (note that their estimates are made for the electrostatic fluctuation spectra, using a Pad\'e approximation for the finite Larmor radius contributions) that calls for a steeper energy spectra slope than the $\sim k^{-4/3}$ power law\cite{Plunk:2010p1360}.

\subsection{Scale fluxes and their angular densities}

\begin{figure}[t]
\begin{center}
\includegraphics[width = 0.48\textwidth]{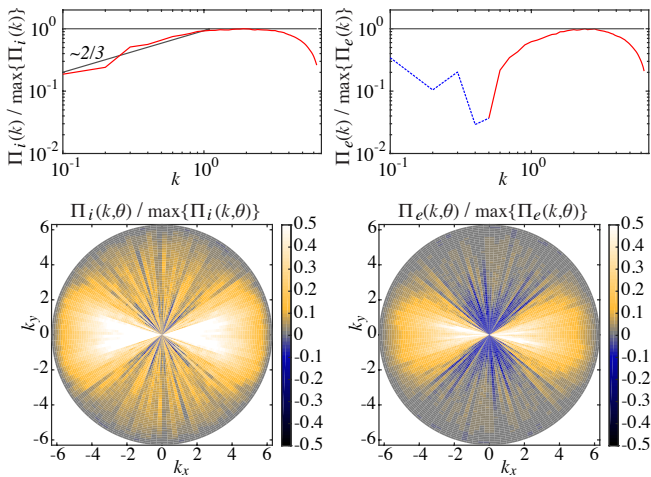}
\end{center}
\caption{(Color online) The free-energy scale fluxes and their respective angular densities for $\omega_{ns}= 1.6$. Temporal averaging is performed in the interval indicated by the full black lines in Fig.~\ref{fig_free-energy}-(a). The dotted blue line indicate negative values for the flux and the normalizations are done in respect to each quantity maximal absolute value.}
\label{fig_fluxes_16}
\end{figure}

\begin{figure}[b]
\begin{center}
\includegraphics[width = 0.48\textwidth]{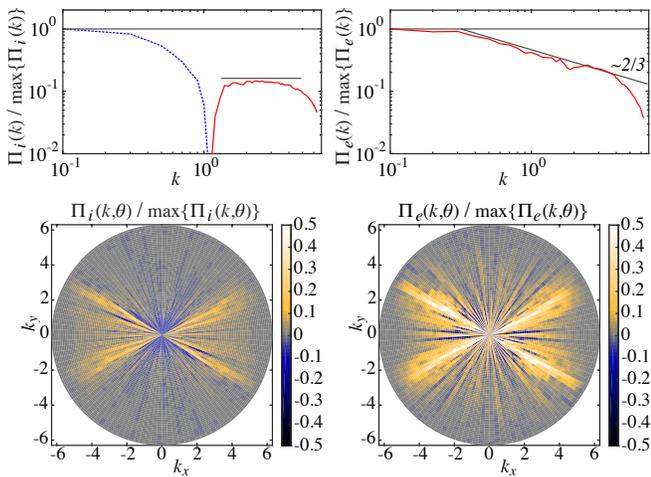}
\end{center}
\caption{(Color online) The free-energy scale fluxes and their respective angular densities for $\omega_{ns}= 4$. Temporal averaging is performed in the interval indicated by the full black lines in Fig.~\ref{fig_free-energy}-(a). The dotted blue line indicate negative values for the flux and the normalizations are done in respect to each quantity maximal absolute value.}
\label{fig_fluxes_40}
\end{figure}

Central to this work, for the steady state regimes, we present the free energy scale flux (Eq.~\ref{def2_f}) and its respective angular density (Eq.~\ref{def2_fa}) for each species. These quantities are presented for $\omega_{ns}= 1.6$ in Fig.~\ref{fig_fluxes_16} and for $\omega_{ns}= 4$ in Fig.~\ref{fig_fluxes_40}. 

Looking at the maximal values, the ion scale flux is about twice as large as the electron flux ($\max\{|\Pi_i(k)|\}\approx 2 \max\{|\Pi_e(k)|\}$) for the $\omega_{ns}= 1.6$ case. At large scales ($k<1$), while the ion flux is slowly growing in value (seemingly as $k^{2/3}$), the electron flux exhibits an inverse cascade process (negative valued flux). At smaller scales ($k>1$), the ion scale flux becomes scale independent, a fact associated with inertial range dynamics. For the $\omega_{ns}= 4$ case, Fig.~\ref{fig_fluxes_40}, the maximal value of the ion scale flux is smaller than its electron counterpart ($\max\{|\Pi_i(k)|\}\approx 0.75 \max\{|\Pi_e(k)|\}$). In this case, it's the ion scale flux that exhibits an inverse cascade process at small wavenumbers ($k<1$). For small scales ($k>1$), the ion scale flux seems to achieve scale independence and is positive in value, even if this value is much smaller that the maximal negative one. A more interesting behavior is exhibited by the electron scale flux, which in this case exhibits a $\sim k^{-2/3}$ scaling. This yet unexplained scaling has been observed for other systems (astrophysical and tokamak conditions) and merits more future work to try to explain it.

While the scale flux represents a good way to determine the type of energy cascade (direct or inverse) and provides additional information regarding its saturation value and wavenumber dependence (or lack thereof), by construction, it cannot distinguish between contributions made by different directions. For both cases and both species, looking at the angular density of the scale flux we clearly see the anisotropic wavenumber makeup. 

In the case of $\omega_{ns}= 1.6$, while negative angular sections are present for the ion flux, the electron angular density clearly shows that both negative and positive angular sections coexist. While the dominance of one aspect gives the overall direct or inverse characteristic to the scale flux, it should be clear that both energy redistribution mechanisms are present at the same time. For the electrons we see that the positive fluxes (from large to small scales) are in the $x$-direction, while the negative fluxes (traditionally associated with the reorganization of large scales in turbulence) are along the $y$-direction. This indicates the radial break up of large structures into smaller ones, while at the same time joining together to form even larger scales in the $y-$direction; a picture consistent with what we come to expect from turbulence in the presence of zonal flows.

For the case $\omega_{ns}= 4$, the positive valued electron scale flux is clearly shown to be composed of negative angular sections, when looking at the angular density. Moreover, we see that both species exhibit dominant positive value fluxes in the $k_x \sim k_y$ directions. We associate this behavior with the vortex dominated turbulent state. As vortices are structures with comparable $k_x$ and $k_y$ wavenumbers, observing preferential $k_x \sim k_y$ directions in the energy redistribution is not surprising. Looking at the ions, we also notice that the contributions to the flux are close to zero along the $k_y$-axis, indicating the suppression of the energy cascade along this direction at larger $k$.

\subsection{The angular flow of energy}

\begin{figure}[b]
\begin{center}
\includegraphics[width = 0.48\textwidth]{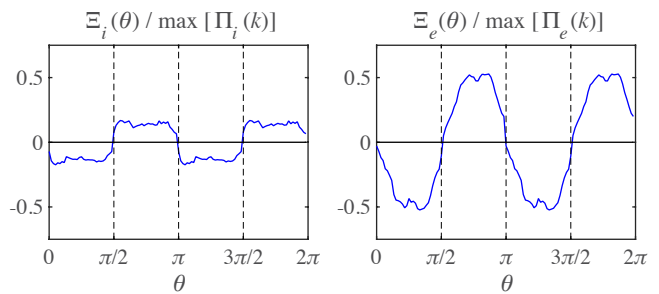}
\end{center}
\caption{(Color online) The angular flux $\Xi(\theta)$, average over the steady state interval for $\omega_{ns}= 1.6$.}
\label{fig_aflux16}
\end{figure}

\begin{figure}[t]
\begin{center}
\includegraphics[width = 0.48\textwidth]{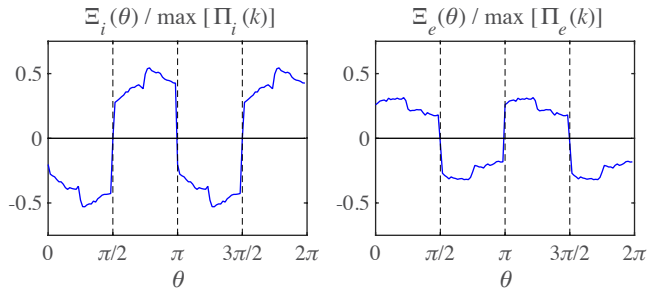}
\end{center}
\caption{(Color online) The angular flux $\Xi(\theta)$, average over the steady state interval for $\omega_{ns}= 4$.}
\label{fig_aflux40}
\end{figure}

While the angular density of the scale flux allows us to measure the anisotropy of the cascade of energy between large and small scales, it does not tell us how much energy is moved from one direction to the other. This is important, as gauging the amount of energy transferred through a direction can help us understand the development of anisotropy. We present the angular flux ($\Xi_s$) for the case $\omega_{ns}= 1.6$ in Fig.~\ref{fig_aflux16} and for the case $\omega_{ns}= 4$ in Fig.~\ref{fig_aflux40}. 

Normalizing the angular flux to the maximal value of the scale flux allows us to assess the strength of the angular energy redistribution. While for an ideal isotropic case we expect to see zero angular flux, oscillations around zero are to be expected in practice, especially when long time averages are not possible.

For $\omega_{ns}= 1.6$, for both ion and electron angular fluxes, negative values can be see in the $[0, \pi/2]$ interval and positive values in the $[\pi/2, \pi]$ interval. This is a sign of energy flowing from the $k_y$-axis towards the $k_x$-axis. The $[\pi, 2\pi]$ interval just reproduces this behavior, due to the reality condition. It is interesting to note that the ion angular flux is substantially reduced in value compared to its flux across k. Also, as the ion angular flux curve is more flat across each interval compared to the electrons, we can infer a more constant angular flow of energy.  

For $\omega_{ns}= 4$, looking at the angular fluxes, we see that the energy flows in opposite directions for ion and electron species. While for the ions the energy is flowing towards the $k_x$-axis, for the electrons the energy is flowing towards the $k_y$-axis. As the angular flux is integrated over all possible scales, we cannot say if the energy represents an exchange between streamer and zonal-flow modes or smaller scale structures.

\section{Discussion and conclusions}

In this paper we introduced a series of novel diagnostics related to the assessment of anisotropy of the nonlinear energy redistribution between scales for turbulent systems. Results are reported for a gyrokinetic plasma in a Z-pinch magnetic geometry. 

From the angular density of the scale flux we see that while scale fluxes indicative of direct energy cascades are observed for GK turbulent systems, they are highly anisotropic. Not only do most positive contributions come from the $\theta=0$ direction ($k_x$-axis), or solely from the $\theta\approx\pi/4$ one, but negative values denoting an inverse exchange of energy is found along the $y$ direction ($\theta=\pi/2$). The Z-pinch case reported here is a perfect example where the backscatter of energy in a direct cascade scenario is found to occupy a distinct wavenumber domain. While the direct cascade is associated with the removal of energy at small scales, the backscatter of energy is associated with the self-organization of structures in turbulence. A similar behavior is expected for the more complex sheared toroidal geometry case.

Looking at the angular density plots for the two cases, we can infer another interesting result. While certain wavenumber domains show the backscatter of energy, this occurs for relatively small $k$. At large $k$, the angular density shows positive values for all angles $\theta$. This seems to indicate a certain universality of the small scales, which just move the free energy to ever smaller scales until they become thermalized. Larger direct numerical simulations are needed to properly validate this interpretation. However, this is consistent with what we expect form turbulence: complex self-organization of larger scales, interacting with universal small scales; especially when zonal flows are present. Relating the $\theta$ angle arc for the domain that exhibits a backscatter of energy to the parameters characterizing the zonal flows is a desired goal for the future. 
Moreover, as the development of coherent structures (vortices) is related to intermittency in turbulence, we see that the angular density plots can capture these effects (the $k_x \sim k_y$ enhanced contributions for the $\omega_{ns}= 4$ case).

One particular line of work that will benefit from this analysis consists in the further development of LES models for gyrokinetics. In essence, LES models attempt to reproduce the correct scale flux value for a given $k$, while neglecting the contributions made to this flux from interactions that involve small scales (not present in the simulation). While LES methods are successful in removing energy accumulation at large wavenumbers, they fail when coupled solely with the largest scales that govern the transport levels in plasma. The fact that LES models depend on a constant, which even when dynamically computed is taken to be positive, can explain this failure. As LES methods were primarily developed to remove energy, they are not designed to take into account inverse cascades (negative scale fluxes), for which a negative value constant is presumably needed. However, as seen by looking at the angular density of the scale flux, both positive and negative angular sections coexist for the same $k$. This implies that computing either a positive or negative constant for the LES method will lead to one of the contributions to the flux to be faulty at very large scales. The results presented in this work suggest the need for a constant that has a $\theta$ dependence for each $k$. 

In general, regardless of the source of anisotropy, a similar $T(k_\perp, k_\parallel)$ object can be constructed for any quadratic nonlinearity, where the parallel and perpendicular wavenumbers are taken in reference to the anisotropic direction. Thus, all diagnostics built on the net energy transfer are general and valid for various turbulent problems. Moreover, as the nonlinear diagnostics presented here are much cheeper from a numerical perspective, it is the hope of the authors that they will be used on a larger scale by the plasma community, helping to accumulate information related to the nonlinear energy exchanges for a wide variety of cases. 

Computing the energy transfers between various wavenumber shells and angular sections is technically possible; this was done in the past for three-dimensional magnetohydrodynamic turbulence using a ring decomposition\cite{Teaca:2009p628}. However, in practice, the computational cost compared to the benefits of such an analysis makes this a prohibitive endeavor for GK turbulence.

In a tokamak geometry, and even for simpler flux-tube simulations, due to the metric dependence on $z$, a $T(k_x, k_y, z)$ object needs to form the starting point for a similar analysis, to allow for the recovery of scale fluxes. While this involves a larger memory requirement (not a particularly stringent constraint), the computational costs remain the same. 
 
Finally, an interesting result that is worth mentioning is the $k^{-2/3}$ slope exhibited by the electron scale flux for steady state turbulence dominated by the electron cascade ($\omega_{ns}=4$). The same $k^{-2/3}$ scaling for the electron scale flux was observed in slab simulations pertinent to astrophysical conditions, while an antenna driven ions species was found to have a scale independent flux. This puzzling and interesting behavior will be analyzed in detail elsewhere and remains an open question.      

\section*{Acknowledgements}
We would like to thank Daniel Told (UCLA) for discussions related to gyrokinetic plasma turbulence in astrophysical conditions. We also acknowledge Paul Crandall (UCLA) for his comments that led to a clearer presentation of the manuscript. The gyrokinetic simulations presented in this work used resources of the National Energy Research Scientific Computing Center, a DOE Office of Science User Facility supported by the Office of Science of the U.S. Department of Energy under Contract No. DE-AC02-05CH11231.


%

\end{document}